\documentclass[prd,english,showpacs,showkeys]{revtex4}%
\usepackage{amsmath}
\usepackage{graphicx}
\usepackage{epsfig}
\usepackage{dcolumn}
\usepackage{bm}
\usepackage{slashed}
\usepackage{amsfonts}
\usepackage{amssymb}
\usepackage{fontenc}
\usepackage{amsmath}
\usepackage{fontenc}
\usepackage{babel}
\usepackage{nameref}%
\providecommand{\U}[1]{\protect\rule{.1in}{.1in}}
 
\begin{document}
\title{Moving point charge as a soliton in nonlinear electrodynamics}
\author{D. M. Gitman$^{1,2,3}$}
\email{gitman@dfn.if.usp.br}
\author{A. E. Shabad$^{1,2}$}
\email{shabad@lpi.ru}
\author{A.A. Shishmarev $^{2,3}$}
\email{aleksei_shishmarev87@mail.ru}
\affiliation{$^{1}$\textsl{P. N. Lebedev Physics Institute, Leninsky Prospekt 53,
Moscow\ 117924, Russia} }
\affiliation{$^{2}$\textsl{Tomsk State University, Lenin Prospekt 36, Tomsk 634050, Russia} }
\affiliation{$^{3}$\textsl{Instituto de F\'{\i}sica, Universidade de S\~{a}o Paulo, CEP
05508-090, S\~{a}o Paulo, S. P., Brazil}}

\begin{abstract}
The field of a moving pointlike charge is determined in nonlinear local
electrodynamics. As a model Lagrangian for the latter we take the one whose
nonlinearity is the Euler-Heisenberg Lagrangian of quantum electrodynamics
truncated at the leading term of its expansion in powers of the first field
invariant. The total energy of the field produced by a point charge is finite
in that model; thereby its field configuration is a soliton. We define a
finite energy-momentum vector of this field configuration to demonstrate that
its components satisfy the standard mechanical relation characteristic of a
free moving massive particle.

\end{abstract}
\maketitle

\section{Introduction}

In this paper we are studying the field of a uniformly moving charge in
nonlinear electrodynamics, in other words we are dealing with nonlinear
extension of the Li\'{e}nard-Wiechert potentials.

The problem of the field caused by a moving charge, besides belonging to
fundamental problems of electrodynamics, is also of certain practical
importance as applied to charged particle beams in accelerators
\cite{experiment}. Its nonlinear extension may have an effect in
small-impact-parameter scattering at high energies where close vicinities of
the charge come into play.

We take the simplest version of nonlinear electrodynamics \cite{CostGitShab},
namely the one that results from keeping only the leading term in the
expansion of the Heisenberg-Euler action \cite{Heisenberg} of quantum
electrodynamics (QED) in powers of the field invariant $\mathfrak{F(}%
x\mathfrak{)=}\frac{1}{4}F_{\mu\nu}F^{\mu\nu}=\frac{1}{2}(B^{2}-E^{2})$, with
the second invariant $\mathfrak{G(}x\mathfrak{)=}\frac{1}{4}F_{\mu\nu
}\widetilde{F}^{\mu\nu}=\mathbf{B\cdot E}$ set equal to zero. (Remind that,
the Heisenberg-Euler action reflects the nonlinearity due to interaction
between electromagnetic fields stemming from their quantum nature, and it is
local in the sense that dependence on space-time derivatives of the fields
present in the full effective action of QED is disregarded from it.) This
nonlinear model may be thought of as an extension of the established theory of
electromagnetism to the close vicinity of a point charge compatible with what
that theory describes at larger distances. When treated seriously, the above
truncated model results in the electric field behavior near the point charge
less singular than the Coulomb law, which provides convergence of the
field-energy integral \cite{CostGitShab}. This means that the field
configuration produced by a point charge may be referred to as a soliton. In
Section \ref{Sec. 2} we present the electric and magnetic fields of this
soliton when it moves with a constant speed. In Section \ref{3} we define the
energy-momentum of this moving soliton, which is just the energy-momentum
vector of a moving particle with its mass defined by an integral of its field.
It is proportional to the charge squared.

The following remark is in order. Our approach to considering the nonlinear
problem includes as its starting point a covariant representation for the
current, the field,\ the energy-momentum tensor and vector. In the trivial,
linear limit, this approach results in totally reproducing the field of a
charge moving with a constant speed as given in the Landau and Lifshitz
text-book \cite{Landau2}. This seemingly undistinguished fact is worth
stressing, however, because, strange though it may seem, one faces a discord
in the present-day literature, many basic text-books included, concerning this
fundamental point (see the analysis in \cite{Field}). The dispute is fueled by
the expression for the field of a moving charge given without a detailed
derivation in Feynman's lectures \cite{Feynman}, that may seem to be different
from that of Ref. \cite{Landau2}, the more so that apparently different
physical ideas are layed into it. This fact even called into being an
experimental verification \cite{experiment} that resulted in favor of
\cite{Landau2}. For this reason we found it not out of place to analyse the
result of \cite{Feynman} in a separate forthcoming publication \cite{6a} to reveal its complete coincidence with
that of \cite{Landau2} both for accelerated motion of the charge and its
motion with a constant speed.

\section{Minimally nonlinear model\label{Sec.2}\qquad}

The Lagrangian of a minimally nonlinear electrodynamics is defined as%
\begin{equation}
L(x)=-\mathfrak{F(}x\mathfrak{)+}\frac{\sigma}{2}\left(  \mathfrak{F(}%
x\mathfrak{)}\right)  ^{2}, \label{lagrangian}%
\end{equation}
where $\mathfrak{F(}x\mathfrak{)}$ makes up (with the reversed sign) the
Lagrangian density of the standard linear Maxwell electrodynamics, while the
second term in (\ref{lagrangian}) is the quartic in the field-strength
addition to it. The field-strength tensor is related to the four-vector
potential $A^{\mu}(x)$ as $F^{\mu\nu}=\partial^{\mu}A^{\nu}(x)-\partial^{\nu
}A^{\mu}(x)$, where $\partial^{\mu}=\frac{\partial}{\partial x_{\mu}}$. Hence,
the first pair of the Maxwell equations, $\left[  \nabla\times\mathbf{B}%
\right]  =0$ and $\left[  \nabla\times\mathbf{E}\right]  +\partial
^{0}\mathbf{B=0}$\textbf{,} with the electric and magnetic field strengths
$E_{i}=F^{i0}$ and $B_{i}=\epsilon_{ijk}F^{jk}$, remains standard. The
self-coupling constant $\sigma$ is presumably small enough. It has the
dimensionality of inverse fourth power of mass \footnote{Here and in what
follows we use standard convention about summation over repeated indices;
Roman letters $i,$ $j,$ $k$ run from 1 to 3, while Greek letters $\mu
,\nu,\lambda$ run from 0 to 3. The three-dimensional vectors are boldfaced.
Their scalar and vector products are defined, respectively, as $\left(
\mathbf{D\cdot C}\right)  =D_{i}C_{i},$ $\left[  \mathbf{D\times C}\right]
_{i}=\epsilon_{ijk}D_{j}C_{k}$.}\emph{.} The four-dimensional scalar product
is $\left(  ux\right)  =u_{0}x_{0}-\left(  \mathbf{u\cdot x}\right)  ,$
$x^{2}=x_{0}^{2}-|\mathbf{x|}^{2}$. We use Lorentz--Heaviside system of
units.\emph{ }If the quartic correction in (\ref{lagrangian}) is understood as
the lowest term in the expansion of the Heisenberg-Euler Lagrangian in powers
of $\mathfrak{F}$, with $\mathfrak{G}=0,$ the value \cite{BerLifPit} of the
photon selfcoupling constant is $\sigma=e^{4}\hbar/(45\pi^{2}m^{4}c^{7})$, where $e$ and
$m$ are electron charge and mass, respectively.

The action includes the selfinteraction of the electromagnetic field and
interaction with a current $j^{\nu}$, produced by a moving charge. It is
\[
S[A]=-\frac{1}{c}\int(\frac{1}{c}A_{\mu}j^{\mu}-L(x))d^{4}x.
\]
The second pair of Maxwell equations should be calculated in assumption that
the trajectory of the charge is fixed, and that its variation, and, thereby,
the variation of the current, are zero. Thus, the variation of the action
takes the form
\begin{align}
\delta S  &  =-\frac{1}{c}\int\left[  \frac{1}{c}j^{\mu}\delta A_{\mu
}+(1\mathfrak{-}\sigma\mathfrak{F(}x\mathfrak{)})\delta\mathfrak{F(}%
x\mathfrak{)}\right]  d\Omega,\nonumber\\
\delta\mathfrak{F(}x)  &  \mathfrak{=}\frac{1}{2}F^{\mu\nu}\delta F_{\mu\nu
}=-\frac{1}{2}F^{\mu\nu}(\frac{\delta\partial A_{\mu}}{\partial x^{\nu}}%
-\frac{\delta\partial A_{\nu}}{\partial x^{\mu}})=-F^{\mu\nu}\frac{\partial
}{\partial x^{\nu}}\delta A_{\mu}. \label{eq8}%
\end{align}
After integrating by parts and taking into account that there is no field at
infinities the latter equation takes the form
\begin{equation}
\delta S=-\frac{1}{c}\int\left[  \frac{1}{c}j^{\mu}+\frac{\partial}{\partial
x^{\nu}}\left[  (1\mathfrak{-}\sigma\mathfrak{F(}x\mathfrak{)})F^{\mu\nu
}\right]  \right]  \delta A_{\mu}d\Omega\label{eq9}%
\end{equation}
As always, variations should turn to zero for any $\delta A_{\mu}$, and we
come to the nonlinear field equation
\begin{equation}
\frac{\partial}{\partial x^{\nu}}\left[  (1\mathfrak{-}\sigma\mathfrak{F(}%
x\mathfrak{)})F^{\mu\nu}\right]  =-\frac{1}{c}j^{\mu}. \label{eq10}%
\end{equation}

\section{\qquad Fields of a moving charge\label{Sec. 2}}

\subsection{Current\label{Current}}

In this paper we restrict ourselves only to those currents $j^{\mu}$ that
correspond to a pointlike or any other charge, which is at rest in the origin
in a certain Lorentz reference frame (the rest frame) and is distributed in a
spherically symmetric and time-independent way in that frame. The charge moves
as a whole with the speed $-\mathbf{v}$\textbf{,} $v<1$, along axis 1,
$v_{i}=\delta_{i1}v$, in the inertial frame that moves relative to the rest
frame with the 4-velocity $u_{\mu}$, which is the unit four-vector, $u^{2}%
=1$,with the components $u_{0}=\gamma$, $u_{i}=\gamma\frac{v_{i}}{c}$, where
$\gamma=\left(  1-v^{2}/c^{2}\right)  ^{-1/2}$. The four-current is%
\begin{equation}
j^{\mu}=u_{\mu}\rho(x), \label{j}%
\end{equation}
where $\rho(x)$ is the Lorentz scalar defined as the time-independent charge
density in its rest frame. This implies that under Lorentz boosts and spacial
rotations only the argument of $\rho\ $is transformed.$\ $

To make sure of the validity of the four-vector representation (\ref{j}) let
us first define the current in the rest frame as $j^{\prime0}(x^{\prime}%
)=c\rho(\mathbf{x}^{\prime})$, $\mathbf{j}^{\prime}(x^{\prime})=0$. All
quantities relating to that frame are marked with primes throughout. Applying
the Lorentz transformation to this current%
\begin{equation}
j_{0}=\gamma\left(  j_{0}^{\prime}+\frac{v}{c}j_{1}^{\prime}\right)  ,\text{
\ }j_{1}=\gamma\left(  j_{1}^{\prime}+\frac{v}{c}j_{0}^{\prime}\right)
,\text{ \ }j_{2}=j_{2}^{\prime},\text{ \ }j_{3}=j_{3}^{\prime},\text{ }
\label{lorcur}%
\end{equation}
with the account of the vanishing of its spacial component in the rest frame
$j_{1}^{\prime}=0$, and applying the inverse transformation to the
coordinates
\begin{equation}
x_{0}^{\prime}=\gamma\left(  x_{0}-\frac{v}{c}x_{1}\right)  ,\text{ \ }%
x_{1}^{\prime}=\gamma\left(  x_{1}-\frac{v}{c}x_{0}\right)  ,\text{ \ }%
x_{2}^{\prime}=x_{2},\text{ \ }x_{3}^{\prime}=x_{3}. \label{inverse_lorentz}%
\end{equation}
we get for the moving charge (\ref{lorcur})$\ j^{0}(x)=\gamma c\rho(\left(
x_{1}\mathbf{-}\frac{v}{c}x_{0}\right)  \gamma,x_{2},x_{3})$, 
$\mathbf{j}(x)=\mathbf{v}\gamma\rho(\left(  x_{1}\mathbf{-}\frac{v}{c}%
x_{0}\right)  \gamma,x_{2},x_{3})$. This agrees with Eq. (\ref{j}). For the
pointlike charge $e$ we have
\begin{align}
j^{\prime0}(x^{\prime})  &  =c\rho(\mathbf{x}^{\prime})=ce\delta^{3}%
(\mathbf{x}^{\prime}),\text{ \ \ \ \ \ \ }\mathbf{j}^{\prime}(x^{\prime
})=0\nonumber\\
j^{0}(x)  &  =\gamma ce\delta(\left(  x_{1}\mathbf{-}\frac{v}{c}x_{0}\right)
\gamma)\delta(x_{2})\delta(x_{3})=ce\delta(x_{1}\mathbf{-}\frac{v}{c}%
x_{0})\delta(x_{2})\delta(x_{3})=\gamma ce\delta^{3}(\mathbf{x}^{\prime
}\mathbf{)}.\nonumber\\
\mathbf{j}(x)  &  =\mathbf{v}\gamma e\delta\left(  \left(
x_{1}\mathbf{-}\frac{v}{c}x_{0}\right)  \gamma\right)  \delta(x_{2}%
)\delta(x_{3})=\mathbf{v} e\delta\left(  x_{1}-\frac{v}{c}%
x_{0}\right)  \delta(x_{2})\delta(x_{3})=\mathbf{v} \gamma
e\delta^{3}(\mathbf{x}^{\prime}). \label{j''}%
\end{align}
Note that our final expressions for the current components differ by the
Lorentz-factor $\gamma$ from the corresponding expressions in Ref.
\cite{Landau2}, because the charge density $\rho(x)$ is defined there as the
zero-component of the four-current, and not as a Lorentz scalar, like here.
The both ways are equivalent, of course.

Later we shall see that the current (\ref{j''}) is reproduced as the
right-hand side of the field equation (\ref{eq10}) for a moving point charge
in nonlinear, as well as linear, electrodynamics, and also in the weak
continuity equation (\ref{partial_conservation}) for the energy-momentum tensor.

\subsection{Covariant ansatz\label{2.2}}

Once the motion of the charge is given, there are no other vectors in the
problem besides $u_{\mu}$ and the four-coordinate of the observation point
$x_{\mu}$, which is the argument of the differential equations (\ref{eq10}).
Therefore, the potential produced by the charge may have the only
representation
\begin{equation}
A^{\mu}=u^{\mu}f_{1}(xu,x^{2})+x^{\mu}f_{2}(xu,x^{2}), \label{potential}%
\end{equation}
where $f_{1}$and $f_{2}$ are functions of the indicated scalars. Then the
field-strength tensor resulting from this potential is%
\begin{equation}
F_{\mu\nu}=(u_{\mu}x_{\nu}-u_{\nu}x_{\mu})f(xu,x^{2}) \label{field tens}%
\end{equation}
with
\begin{equation}
f(xu,x^{2})=\frac{\partial f_{2}}{\partial(xu)}-2\frac{\partial f_{1}%
}{\partial(x^{2})}.
\end{equation}
The requirement that in the rest frame the (electric) field $F_{0i}$\ be
independent of time $x_{0}$ implies that $f$ should be a function of the
combination
\begin{equation}
W^{2}=\left(  xu\right)  ^{2}-x^{2} \label{W}%
\end{equation}
of its argument. So, by definition, the invariant $W$ is the distance from the
observation point to the charge in the rest frame of the latter $W=|\mathbf{x}%
^{\prime}|=r$. (In the rest frame it is evident that Eq. (\ref{W}) defines a
positive quantity $W=|\mathbf{x}^{\prime}|=r$. As long as $W^{2}$ is a Lorentz
invariant it remains positive in any frame.) Therefore, in what follows we
shall keep to the representation%
\begin{equation}
f(xu,x^{2})=g\left(  W^{2}\right)  . \label{g}%
\end{equation}
It can be directly verified that the field tensor resulting from
(\ref{field tens}) and (\ref{g})
\begin{equation}
F_{\mu\nu}=(u_{\mu}x_{\nu}-u_{\nu}x_{\mu})g\left(  W^{2}\right)
\label{field-of-Z}%
\end{equation}
satisfies the first pair of the Maxwell equations (the Bianchi identities)
\begin{equation}
\epsilon^{\mu\nu\lambda\rho}\frac{\partial F_{\nu\lambda}}{\partial x^{\rho}%
}=0,\text{ \ }\epsilon^{0123}=1. \label{Bianchi}%
\end{equation}
However, it is easier to see this if we note that by choosing $\ f_{2\text{ }%
}=0$ in (\ref{potential}) and taking $f_{1\text{ }}$as%

\[
f_{1}(W^{2})=\frac{1}{2}\int g\left(  W^{2}\right)  \text{d}W^{2}%
\]
we determine the vector-potential generating the field (\ref{field-of-Z}),
(\ref{g})
\[
A_{\mu}=u_{\mu}f_{1}(W^{2})
\]
thus guaranteeing the fulfillment of (\ref{Bianchi}). In the rest frame this
potential is the 3-scalar $A_{0}^{\prime}=$ $f_{1}(\mathbf{x}^{\prime2})$,
$\mathbf{A}^{\prime}=0$. Setting $f_{2\text{ }}=0$ is not the only way to fix
the vector potential generating Eq. (\ref{field-of-Z}). Another choice of the
potential admitted within the gauge arbitrariness may be, for instance,
\[
A_{\mu}=\left(  u_{\mu}\left(  ux\right)  -x_{\mu}\right)  \left(  ux\right)
g(W^{2}).
\]
This potential satisfies the Lorentz-invariant gauge condition $\left(
Au\right)  =0$.

In what follows we concentrate in finding a solution to the second pair of the
nonlinear Maxwell equations (\ref{eq10}) using the form (\ref{field-of-Z}) as
an ansatz for searching for it.

\subsubsection{Linear limit\label{2.2.1}}

Let us first note that the covariant extension of the Coulomb field produced
by a moving point charge $e$, which is at rest in the origin in the rest
frame, i.e., by the one, whose world line passes through the point
$\widetilde{x}_{0}=\widetilde{x}_{i}=0$, is:
\begin{equation}
F_{\mu\nu}^{\text{lin}}=\frac{(u_{\mu}x_{\nu}-u_{\nu}x_{\mu})}{W}\frac{e}{4\pi
W^{2}}. \label{covariant_F}%
\end{equation}
This corresponds to setting
\begin{equation}
g^{\text{lin}}(W^{2})=\frac{e}{4\pi W^{3}} \label{glin}%
\end{equation}
in (\ref{field-of-Z}) in the linear limit. In the rest frame $u_{i}=0$,
$u_{0}=1$, we have $W=|\mathbf{x}^{\prime}|=r$, and then the nonvanishing
components of Eq.(\ref{covariant_F}) constitute the Coulomb electric field of
a point-like charge:
\begin{equation}
F_{0i}^{\prime\text{lin}}=\frac{x_{i}^{\prime}}{r}\frac{e}{4\pi r^{2}}.
\end{equation}
Expression (\ref{covariant_F}) satisfies, in the moving frame, the linear
Maxwell equation%
\begin{equation}
\partial^{\mu}F_{\mu\nu}^{\text{lin}}=\frac{1}{c}j_{\nu} \label{lineq}%
\end{equation}
\emph{ }with the current (\ref{j''}) of a point charge. This rather evident
statement is explicitly demonstrated in \ref{Ap1}\emph{.}

In order to reproduce the standard form of the electric and magnetic field
components produced by a charge moving with the time-independent speed
$\overset{\cdot}{u}_{\mu}=0$ with its worldline $\mathbf{\tilde{x}}%
=\frac{\mathbf{v}}{c}\widetilde{x}_{0}$ passing through the origin
$\mathbf{\tilde{x}}=0$ at zero time-moment $\widetilde{x}_{0}=0$, written in
Ref. \cite{Landau2} in the case of linear Maxwell electrodynamics as
\begin{equation}
\mathbf{E}^{\text{lin}}=(1-\frac{v^{2}}{c^{2}})\frac{e(\mathbf{x-}%
\frac{\mathbf{v}}{c}x_{0})}{4\pi R^{\ast3}},\text{ \ }\mathbf{B}^{\text{lin}%
}\mathbf{=[v\times E}^{\text{lin}}\mathbf{],} \label{E,B}%
\end{equation}
it is sufficient to note that $R^{\ast}$ defined in \cite{Landau2} (when
$\mathbf{v}$ has only the first component, $v_{i}=\delta_{i1}v$) as%

\begin{equation}
R^{\ast2}=(x_{1}\mathbf{-}\frac{v}{c}x_{0})^{2}+\left(  x_{2}^{2}+x_{3}%
^{2}\right)  \left(  1-\frac{v^{2}}{c^{2}}\right)  \label{R*}%
\end{equation}
is related to the Lorentz scalar $W$ (\ref{W}) in the following way $R^{\ast
2}=\left(  1-\frac{v^{2}}{c^{2}}\right)  W^{2}$ (Note that $R^{\prime}$ of
Ref. \cite{Landau2} is just our invariant $W$). With this substitution Eq.
(\ref{E,B}) is just what follows for the electric and magnetic components of
(\ref{covariant_F}).

The electric and magnetic fields (\ref{E,B}) or (\ref{covariant_F}) can be
written also as functions of the difference\emph{ }$\Delta x_{\mu}=x_{\mu
}-\widetilde{x}_{\mu}$\emph{ }between the coordinate of the observation
point\emph{ }$x_{\mu}$\emph{ }and that of the charge\emph{ }$\widetilde{x}%
_{\mu}$\emph{ }as follows\emph{ }%
\begin{align}
E_{i}^{\text{lin}}(\Delta x,\tilde{x})  &  =F^{\text{lin}0i}=\frac{e}{4\pi
W^{2}}\frac{(u^{0}\Delta x^{i}-u^{i}\Delta x^{0})}{W},\nonumber\\
B_{i}^{\text{lin}}(\Delta x,\tilde{x})  &  =\epsilon_{ijk}F_{jk}^{\text{lin}%
}=\epsilon_{ijk}\frac{e}{4\pi W^{2}}\frac{\left(  u_{j}\Delta x_{k}%
-u_{k}\Delta x_{j}\right)  }{W} \label{gener_expression}%
\end{align}
with%
\begin{equation}
W^{2}=\left(  1-\frac{v^{2}}{c^{2}}\right)  ^{-1}\left[  (\Delta
x_{1}\mathbf{-}\frac{v}{c}\Delta x_{0})^{2}+\left(  \left(  \Delta
x_{2}\right)  ^{2}+\left(  \Delta x_{3}\right)  ^{2}\right)  \left(
1-\frac{v^{2}}{c^{2}}\right)  \right]  \label{W1}%
\end{equation}
As long as the trajectory is fixed, neither\emph{ }(\ref{E,B}), nor
(\ref{gener_expression}) contain the variable\emph{ }$\widetilde{x}_{\mu}%
$\emph{ }explicitly.\emph{ }

In order to write the fields in the Li\'{e}nard-Wiechert form it is necessary
to exploit the light-cone condition\emph{ }$\left(  \Delta x\right)  ^{2}=0$,
i.e.\emph{ }$\left(  \Delta x_{2}\right)  ^{2}+\left(  \Delta x_{3}\right)
^{2}=\left(  \Delta x_{0}\right)  ^{2}-\left(  \Delta x_{1}\right)  ^{2}%
$\emph{ }in (\ref{gener_expression}), which tells that the charge and the
observation point must be separated by a light-like interval for the field
produced by the charge be nonzero in the observation point. Imposing the
light-cone condition turns (\ref{W1}) into\emph{ }%
\[
W^{2}=\left(  1-\frac{v^{2}}{c^{2}}\right)  ^{-1}\left(  \Delta x_{0}-\frac
{v}{c}\Delta x_{1}\right)  ^{2},
\]
and (\ref{gener_expression}) becomes Eqs. (63,8), (63,9) of Ref.
\cite{Landau2} with\emph{ }$\overset{\cdot}{\mathbf{v}}=0$\emph{ }after the
identification\emph{ }$R=\Delta x_{0}$,\emph{ }$R_{i}=\Delta x_{i}$.

In this connection, the following remark is in order. In linear
electrodynamics the influence of a charge propagates with the speed of light
in the vacuum $c$, and hence the observation point must be separated from the
four-position of the source by a light-like interval, $\left(  \Delta
x\right)  ^{2}=0$, other space-time points carrying no field produced by the
source at this position. This is not the case in nonlinear electrodynamics. It
is well known already in QED that the nonlinearity leads \cite{Bialin} to
nontrivial dielectric permeability and magnetic susceptibility of the vacuum
in an external field, thereby to deviation of the speed of propagation from
that of light. The role of the external field is in our case played by the
field produced by the charge itself, the propagation speed depending, as a
matter of fact, on its intensity.\emph{ }For this reason it will be not
relevant to impose the light-cone condition onto the nonlinear solution of the
next section for getting an analog of the Li\'{e}nard-Wiechert
potential.\qquad

\subsection{Solution to nonlinear field equations}

From (\ref{eq10}) we can conclude that its solution may be related to
$F^{\text{lin}\mu\nu}$ given by (\ref{covariant_F}) as
\begin{equation}
(1-\sigma\mathfrak{F}(x))F^{\mu\nu}=F^{\text{lin}\mu\nu},
\label{tensor_connection}%
\end{equation}
since (\ref{covariant_F}) satisfies the linear Maxwell equation (\ref{lineq})
as demonstrated in \ref{Ap1}. Then, using the ansatz (\ref{field-of-Z}) and
taking into account that $u^{2}=1$, we get
\begin{align}
&  (1-\sigma\mathfrak{F}(x))g=\frac{e}{4\pi W^{3}},\nonumber\\
&  \mathfrak{F}(x)=\frac{1}{4}F_{\mu\nu}F^{\mu\nu}=-\frac{1}{2}W^{2}g^{2}.
\label{connection}%
\end{align}
This is the cubic equation for $g(W^{2})$:%
\begin{equation}
g^{3}+\frac{2g}{\sigma W^{2}}-\frac{2e}{4\pi\sigma W^{5}}=0. \label{cubic}%
\end{equation}
The discriminant of this equation is positive%
\begin{equation}
Q=\left(  \frac{2}{3\sigma W^{2}}\right)  ^{3}+\left(  \frac{e}{4\pi\sigma
W^{5}}\right)  ^{2}>0, \label{determinant}%
\end{equation}
because $W^{2}>0$, as argued before. The only real Cardano solution for the
case is%
\begin{equation}
g=\sqrt[3]{\frac{e}{4\pi\sigma W^{5}}+\sqrt{Q}}+\sqrt[3]{\frac{e}{4\pi\sigma
W^{5}}-\sqrt{Q}}. \label{solution}%
\end{equation}
Its asymptotic behavior near the charge%
\begin{equation}
g\sim\left(  \frac{2e}{4\pi\sigma W^{5}}\right)  ^{1/3}\ { \ }\text{as}{
\ }W\rightarrow0, \label{asymp}%
\end{equation}
will provide convergence for the field mass integral (\ref{M_}) in Subsection
"Field mass". In the rest frame, when substituted into (\ref{field-of-Z}) with
$u_{0}=1$, $u_{i}=0$, $W=r$ this reproduces the result of Ref.
\cite{CostGitShab}. In the limit of vanishing nonlinearity $\sigma
\rightarrow0$ Eq. (\ref{solution}) becomes $g=\frac{e}{4\pi W^{3}}$, therefore
the solution Eq. (\ref{field-of-Z}) turns into expression \ (\ref{covariant_F}%
) for the known field of a moving charge in the linear Maxwell
electrodynamics, $F_{\mu\nu}=F_{\mu\nu}^{\text{lin}}$.

Finally, from (\ref{field-of-Z}) for the point charge moving along the
trajectory $\mathbf{\tilde{x}}=\frac{v}{c}\tilde{x}_{0}$ with a constant speed
along axis 1, expressions for the electric and magnetic fields $E_{i}=F^{0i}$
and $B_{i}=\epsilon_{ijk}F_{jk}$ as functions of the observation coordinates
are
\begin{equation}
\mathbf{E}=(1-\frac{v^{2}}{c^{2}})^{-1/2}(\mathbf{x-}\frac{\mathbf{v}}{c}%
x_{0})g(W^{2}),\text{ \ }\mathbf{B=[v\times E].} \label{observation}%
\end{equation}
In terms of the distance from the charge $\Delta x_{\mu}=x_{\mu}-\tilde
{x}_{\mu}$ these are%
\begin{align}
\mathbf{E}  &  =\left(  1-\frac{v^{2}}{c^{2}}\right)  ^{-1/2}g(W^{2})\left(
\Delta\mathbf{x}-\frac{\mathbf{v}}{c}\Delta x_{0}\right)  ,\nonumber\\
B_{1}  &  =0,\text{ \ }B_{2}=-2\frac{v}{c}\left(  1-\frac{v^{2}}{c^{2}%
}\right)  ^{-1/2}g(W^{2})\Delta x_{3},\text{ \ }B_{3}(\Delta x)=2\frac{v}%
{c}\left(  1-\frac{v^{2}}{c^{2}}\right)  ^{-1/2}g(W^{2})\Delta x_{2}.
\label{finally}%
\end{align}
where $W^{2}$ is the same as (\ref{W1}), and $g(W^{2})$ is the solution
(\ref{solution}). This is the nonlinear generalization of Eqs. (\ref{E,B}) and
(\ref{gener_expression}).

\section{Solitonic representation of the point-charge field\label{3}}

\subsection{Energy-momentum tensor\label{3.1}}

In this section we set $c=1$, and also $\hbar=1$ in the expression for $\sigma$. The Noether current corresponding to space-time translations
\[
T^{\mu\nu}=\frac{\partial A^{\lambda}}{\partial x_{\mu}}\frac{\partial
L(x)}{\partial\left(  \partial A^{\lambda}/\partial x_{\nu}\right)  }%
-g^{\mu\nu}L(x),
\]
calculated using the Lagrange density (\ref{lagrangian}) with the account of
the field equations (\ref{eq10}) is written as
\begin{equation}
T^{\mu\nu}=-\left[  1-\sigma\mathfrak{F(}x\mathfrak{)}\right]  F^{\mu\lambda
}F_{\lambda}^{\nu}-\frac{\partial}{\partial x_{\lambda}}\left[  A^{\mu}\left(
1-\sigma\mathfrak{F(}x\mathfrak{)}\right)  F_{\lambda}^{\nu}\right]
-g^{\mu\nu}L(x). \label{en_stress_tensor2}%
\end{equation}
This is the energy-momentum tensor of the electromagnetic field. The
requirement that it should be gauge-invariant makes us omit the
potential-dependent second term (that is a full derivative) to be left with
the expression, symmetrical under the permutation of the indices:
\begin{equation}
T^{\mu\nu}=-\left[  1-\sigma\mathfrak{F(}x\mathfrak{)}\right]  F^{\mu\lambda
}F_{\lambda}^{\nu}-g^{\mu\nu}L(x). \label{T}%
\end{equation}
Densities of the energy $T^{00}$ and of the Pointing vector $T^{i0}$ have the
form%
\begin{align}
T^{00}  &  =-F^{0\lambda}F_{\lambda}^{0}\left(  1-\sigma\mathfrak{F(}%
x\mathfrak{)}\right)  +\mathfrak{F(}x\mathfrak{)-}\frac{\mathfrak{\sigma}}%
{2}\left(  \mathfrak{F(}x\mathfrak{)}\right)  ^{2},\nonumber\\
T^{i0}  &  =-F^{i\lambda}F_{\lambda}^{0}\left(  1-\sigma\mathfrak{F(}%
x\mathfrak{)}\right)  . \label{energy_pointing}%
\end{align}
Contervariant components of the tensor (\ref{T}) are
\begin{equation}
T_{v}^{\mu}=-\left[  1-\sigma\mathfrak{F(}x\mathfrak{)}\right]  F^{\mu\lambda
}F_{v\lambda}-\delta_{v}^{\mu}L(x).
\end{equation}
Its trace is different from zero%
\begin{equation}
T_{i}^{i}=2\sigma\left(  \mathfrak{F(}x\mathfrak{)}\right)  ^{2}\mathfrak{.}
\label{trace}%
\end{equation}
in the nonlinear case $\sigma\neq0$.

\subsubsection{Weak continuity and energy-momentum conservation\label{3.1.1}}

The above construction of the energy-momentum tensor does not provide the
fulfillment of its continuity, since the current is meant to be supported by
outer forces and therefore we should admit the energy-momentum
non-conservation. Instead of the continuity equation we formally obtain what
may be referred to as the "partial conservation of the Noether current" in the
form%
\begin{equation}
\frac{\partial}{\partial x^{\mu}}T_{\nu}^{\mu}=-F_{\nu\lambda
}j^{\lambda}. \label{partial_conservation}%
\end{equation}
To derive this relation, equation of motion (\ref{eq10}) was used in Appendix
2 together with the Bianchi identities $\frac{\partial F_{\lambda\rho
}}{\partial x^{\nu}}=-\frac{\partial F_{\rho\nu}}{\partial x^{\lambda}}%
-\frac{\partial F_{\nu\lambda}}{\partial x^{\rho}}$. This relation is not
owing to the nonlinearity and retains its form in the standard linear theory
as well. The right-hand side in (\ref{partial_conservation}) is not vanishing. However
the continuity of the energy-momentum tensor holds in a weak form%
\begin{equation}
u^{\nu}\frac{\partial}{\partial x^{\mu}}T_{\nu}^{\mu}=0. \label{weak}%
\end{equation}
To see this, note that the 4-current of a point charge moving with the
constant speed is parallel to its 4-velocity vector. Then the right-hand side
of (\ref{partial_conservation}) disappears when contracted with the
4-velocity, $u^{\nu}F_{\nu\lambda}j^{\lambda}\sim u^{\nu}F_{\nu\lambda
}u^{\lambda}=0$. The property of weak continuity (\ref{weak}) will be
sufficient for establishing the conservation of the energy-momentum vector.

Let us define the latter by the integral over a space-like hyperplane that is
orthogonal to the four-velocity and crosses the time axis at $x_{0}%
=s\sqrt{1-v^{2}}$%
\begin{equation}
P_{\mu}=\int u^{\nu}T_{\nu}^{\mu}\delta\left(  ux-s\right)  \text{d}^{4}x.
\label{P}%
\end{equation}
Via the Gauss theorem it follows from the vanishing of the 4-divergence
(\ref{weak}) that $P_{\mu}$ is independent of $s$, and thereby of the time of
observation $x_{0}$, because the hyperplane can be shifted as a whole along
the vector $u^{\nu}$ without affecting the value of the integral, since the
fields (\ref{field-of-Z}) decrease at space-time infinity no less fast as in
the linear electrodynamics, the nonlinearity fades away far from the charge,
where its fields are weak. On the contrary, one cannot change to a space-like
hyperplane inclined differently than in (\ref{P}) in the definition of the
energy-momentum 4-vector due to the lack of the continuity law analogous to
(\ref{weak}) with the unit vector $u^{\nu}$ other than the 4-velocity.

When the field $F_{\nu\lambda}$, on which the energy-momentum tensor $T_{\nu
}^{\mu}$ depends, is that of a uniformly moving (or resting) point charge, the
integral in (\ref{P}) usually diverges and hence makes no sense. This is not
the case in the nonlinear theory under consideration here, as we shall see in
the next subsection. Therefore, we may treat the energy-momentum vector
(\ref{P}) seriously.

Bearing in mind that $s$ is a Lorentz scalar (moreover, set equal to zero in
what follows), and that $T_{\nu}^{\mu}$ (\ref{T}) is a tensor,\ the integral
(\ref{P}) does define a Minkowski vector. It cannot help being directed along
$u^{\nu}$, since this is the only external vector in the integrand of
(\ref{P}). Hence we write%
\begin{equation}
P^{\mu}=u^{\mu}M_{\text{f}}, \label{P1}%
\end{equation}
where $M_{\text{f}}$ is the field mass.

\subsection{Finite field mass\label{3.2}}

It follows from (\ref{P}) that
\[
M_{\text{f}}=P^{\nu}u_{\nu}=\int u_{\nu}T^{\mu\nu}u_{\mu}\delta\left(
ux\right)  \text{d}^{4}x.
\]
From (\ref{T}) and (\ref{covariant_F}) we calculate
\begin{align}
M_{\text{f}}  &  =\int\left[  -\left[  1-\sigma\mathfrak{F(}x\mathfrak{)}%
\right]  u_{\mu}F^{\mu\lambda}F_{\lambda}^{\nu}u_{\nu}-L\right]  \delta\left(
ux\right)  \text{d}^{4}x=\nonumber\\
&  =\int\left[  \left(  1-\sigma\frac{W^{2}g^{2}}{2}\right)  W^{2}g^{2}%
+\frac{W^{2}g^{2}}{2}+\frac{\sigma W^{4}g^{4}}{8}\right]  \delta\left(
ux\right)  \text{d}^{4}x. \label{M_f}%
\end{align}
Thanks to the delta-function we may set $W^{2}=-x^{2}$ in the integrand, the
argument of the function $g(W^{2})$ included. For any function $\Phi(x^{2})$,
provided the integrals below converge, the following chain of relations holds%
\[
\int\Phi(x^{2})\delta\left(  ux\right)  \text{d}^{4}x=\frac{1}{u_{0}}\int%
\Phi(-x_{1}^{2}\frac{u_{0}^{2}-u_{1}^{2}}{u_{0}^{2}}-x_{2}^{2}-x_{3}%
^{2})\text{d}x_{1}\text{d}x_{2}\text{d}x_{3}.
\]
Here we have integrated over $x_{0}$ using the delta-function. Performing the
change of the variable $x_{1}\frac{\sqrt{u_{0}^{2}-u_{1}^{2}}}{u_{0}}%
=x_{1}\sqrt{1-v^{2}}=x_{1}^{\prime}$ and omitting the prime afterwards
we find that this integral is equal to
\[
\int\Phi(-\mathbf{x}^{2})\text{d}^{3}x.
\]
Applying this result to (\ref{M_f}) we get for the field mass the integral
over the space%
\[
M_{\text{f}}=\int\left[  \frac{\mathbf{x}^{2}g\left(  \mathbf{x}^{2}\right)
^{2}}{2}+\frac{3}{8}\sigma\mathbf{x}^{4}g\left(  \mathbf{x}^{2}\right)
^{4}\right]  \text{d}^{3}x.
\]
Note that the function $g\left(  \mathbf{x}^{2}\right)  $ involved here is
just (\ref{solution}) taken in the rest frame.

With the help of Eqs.(\ref{cubic}) the second term can be expressed as%
\[
\frac{3}{8}\sigma\mathbf{x}^{4}g\left(  \mathbf{x}^{2}\right)  ^{4}=\frac
{3}{4}\left(  -\mathbf{x}^{2}g\left(  \mathbf{x}^{2}\right)  ^{2}%
+\frac{eg\left(  \mathbf{x}^{2}\right)  }{4\pi|\mathbf{x|}}\right)  .
\]
Then we can rewrite expression for the mass as
\begin{equation}
M_{\text{f}}=\int\left[  -\frac{\mathbf{x}^{2}g\left(  \mathbf{x}^{2}\right)
^{2}}{4}+\frac{3}{4}\frac{eg\left(  \mathbf{x}^{2}\right)  }{4\pi|\mathbf{x|}%
}\right]  \text{d}^{3}x. \label{M_}%
\end{equation}
Therefore, to calculate the mass $M_{\text{f}}$, we have to calculate two
integrals. The first one is%
\begin{align}
&  \int\mathbf{x}^{2}g\left(  \mathbf{x}^{2}\right)  ^{2}\text{d}%
^{3}x=|e|^{\frac{3}{2}}\left(  \frac{3}{2\sigma(4\pi)^{2}}\right)  ^{\frac
{1}{4}}\frac{3}{2}I_{1},\nonumber\\
&  I_{1}=\int\limits_{0}^{\infty}y^{\frac{2}{3}}\left(  \sqrt[3]{\sqrt
{1+y^{4}}+1}-\sqrt[3]{\sqrt{1+y^{4}}-1}\right)  ^{2}dy=0.885,
\end{align}
and the second one is%
\begin{align}
&  e\int\frac{g\left(  \mathbf{x}^{2}\right)  }{4\pi|\mathbf{x|}}\text{d}%
^{3}x=|e|^{\frac{3}{2}}\left(  \frac{3}{2\sigma(4\pi)^{2}}\right)  ^{\frac
{1}{4}}I_{2,}\nonumber\\
&  I_{2}=\int\limits_{0}^{\infty}y^{-\frac{2}{3}}\left(  \sqrt[3]%
{\sqrt{1+y^{4}}+1}-\sqrt[3]{\sqrt{1+y^{4}}-1}\right)  ^{2}dy=3.984.
\end{align}

The final result is%
\begin{equation}
M_{\text{f}}=|e|^{\frac{3}{2}}\left(  \frac{3}{2\sigma(4\pi)^{2}}\right)
^{\frac{1}{4}}\frac{1}{4}\left(  3I_{2}-\frac{3}{2}I_{1}\right)
=2.65|e|^{\frac{3}{2}}\left(  \frac{3}{2\sigma(4\pi)^{2}}\right)  ^{\frac
{1}{4}}<\infty,\nonumber
\end{equation}
and we can see that this is the same value as the full electrostatic energy of
a stationary pointlike particle found in \cite{CostGitShab}.

\section{Conclusion\label{4}}

The object of the study in this paper is a nonlinear electrodynamics. We
represented a moving pointlike electric charge as a soliton, a particle-like
solution for its electromagnetic field that makes up a field configuration
with finite energy. The property of finiteness of the field mass is inherent
in many nonlinear models, for example in ones in Refs. \cite{Kruglov}, and in
the first place, in the famous Born-Infeld model \cite{BornInfeld}. This
property has been encouraging attempts to attribute the experimental value of
the electron mass to the energy carried by its field, the electromagnetic
contribution being sometimes almost exhausting \cite{note}.

To specialize our consideration, we have chosen the model of Ref.
\cite{CostGitShab}. This nonlinear model originates from truncation of quantum
electrodynamics at the second power of the field invariant $\mathfrak{F,}$and
it may be thought of as the simplest model among those that result in
finiteness of the field energy of a point charge, because it admits analytical
solution to the field equations. Besides, its Lagrangian (\ref{lagrangian})
does not possess the disadvantage of being a singular function of the field.
It is this quality that led to presence of a maximum value of electric field,
responsible, in the end, for the finiteness of the field energy in the
Born-Infeld \cite{BornInfeld}, and other \cite{Kruglov}, \cite{note}, like
models. On the opposite, in the model of \cite{CostGitShab} the field of a
point charge is still singular near the charge (\ref{asymp}), but this
singularity is suppressed to the extent sufficient for the convergence of the
field-mass integral (\ref{M_f}).

By explicitly solving the nonlinear Maxwell equation (\ref{eq10}) in an
inertial Lorentz reference frame with the use of covariant ansatz
(\ref{field-of-Z}) that includes the 4-vector $u_{\mu}$ of the particle speed,
we found the electric and magnetic fields of a uniformly moving charge as
functions of the observation point (\ref{observation}) and of the 4-distance
to the moving charge (\ref{finally}).

We defined a gauge-invariant symmetric energy-momentum tensor (\ref{T}) for
the field of a moving charge. Irrespective of a special choice of nonlinear
model and already in the linear limit $\sigma=0$, the energy-momentum tensor
possesses but a weak conservation property (\ref{partial_conservation}),
(\ref{weak}), because the current, corresponding to the charge moving with a
constant speed is external. We were able to establish, nevertheless, the
time-independence of the energy and momentum as components of the Minkowskian
vector defined as an integral (\ref{P}) of the energy-momentum tensor over a
hyperplane orthogonal to $u_{\mu}$. The energy-momentum vector (\ref{P1})\ is
the same as the mechanical energy-momentum of a mere massive particle
(soliton) with the mass equal to the finite rest energy of the electric field.

The consideration in the present article can be readily extended to include
any local nonlinear electrodynamics with convergent field energy. This
condition is met if the corresponding Lagrangian in place of (\ref{lagrangian}%
) grows with the field invariant already as $\mathfrak{F}^{w}$ with
$w>\frac{3}{2}$ (to be published elsewhere).

\section*{Acknowledgements}

Supported by FAPESP under grants 2013/00840-9, 2013/16592-4 and 2014/08970-1,
by RFBR under Project 15-02-00293a, and by the TSU Competitiveness Improvement
Program, by a grant from \textquotedblleft The Tomsk State University D.I.
Mendeleev Foundation Program\textquotedblright. A.A.S. thanks also CAPES for
support. A.E.S. thanks the University of S\~{a}o Paulo for hospitality
extended to him during the period when this work was being performed.

\section*{Appendix 1\label{Ap1}}

Here we demonstrate that Eq.(\ref{covariant_F}) satisfies linear equation
(\ref{lineq}) with the point-like charge current (\ref{j''}). We omit the
superscript "lin" within this Appendix.

It is convenient to rewrite (\ref{covariant_F}) in terms of coordinates of the
coordinate system $K^{\prime}\,$, which moves along the axis $x_{1}$ with
constant velocity $v$ (rest frame of the moving particle). The corresponding
Lorentz boost is,
\[
x_{0}=\gamma\left(  x_{0}^{\prime}+\frac{v}{c}x_{1}^{\prime}\right)  ,\text{
\ }x_{1}=\gamma\left(  x_{1}^{\prime}+\frac{v}{c}x_{0}^{\prime}\right)
,\text{ \ }x_{2}^{\prime}=x_{2},\text{ \ }x_{3}^{\prime}=x_{3}.
\]
and the inverse transformation is given by (\ref{inverse_lorentz}). The
Lorentz transformation reduces the scalar $W^{2}$ to the form%
\begin{gather}
W^{2}=(xu)^{2}-x^{2}=\nonumber\\
=x_{1}^{\prime2}+x_{2}^{\prime2}+x_{3}^{\prime2}=r^{\prime2}. \label{A2_1}%
\end{gather}
The relations%
\begin{align}
\left.  \frac{\partial x_{0}^{\prime}}{\partial x_{0}}\right\vert _{x}  &
=\gamma\left.  \frac{\partial\left(  x_{0}-\frac{v}{c}x_{1}\right)  }{\partial
x_{0}}\right\vert _{x}=\gamma,\nonumber\\
\left.  \frac{\partial x_{1}^{\prime}}{\partial x_{0}}\right\vert _{x}  &
=\gamma\left.  \frac{\partial\left(  x_{1}-\frac{v}{c}x_{0}\right)  }{\partial
x_{0}}\right\vert _{x}=-\gamma\frac{v}{c},\nonumber\\
\left.  \frac{\partial x_{0}^{\prime}}{\partial x_{1}}\right\vert _{x}  &
=\gamma\left.  \frac{\partial\left(  x_{0}-\frac{v}{c}x_{1}\right)  }{\partial
x_{1}}\right\vert _{x}=-\gamma\frac{v}{c},\nonumber\\
\left.  \frac{\partial x_{1}^{\prime}}{\partial x_{1}}\right\vert _{x}  &
=\gamma\left.  \frac{\partial\left(  x_{1}-\frac{v}{c}x_{0}\right)  }{\partial
x_{1}}\right\vert _{x}=\gamma, \label{A2_2}%
\end{align}
where $\left.  {}\right\vert _{x}$ designates derivative calculated at
constant $x$, follow from (\ref{inverse_lorentz}). Then
\begin{align}
\left.  \frac{\partial}{\partial x_{0}}\right\vert _{x}  &  =\left(
\gamma\left.  \frac{\partial}{\partial x_{0}^{\prime}}\right\vert _{x^{\prime
}}-\gamma\frac{v}{c}\left.  \frac{\partial}{\partial x_{1}^{\prime}%
}\right\vert _{x^{\prime}}\right)  ,\nonumber\\
\left.  \frac{\partial}{\partial x_{1}}\right\vert _{x}  &  =\left(
-\gamma\frac{v}{c}\left.  \frac{\partial}{\partial x_{0}^{\prime}}\right\vert
_{x^{\prime}}+\gamma\left.  \frac{\partial}{\partial x_{1}^{\prime}%
}\right\vert _{x^{\prime}}\right)  . \label{A2_3}%
\end{align}

Now, with the account of (\ref{A2_1})-(\ref{A2_3}), one can calculate
divergency of $F_{\mu\nu}$ exploiting the covariant form (\ref{covariant_F}).
The zeroth and first components are:%

\begin{align}
\partial^{\mu}F_{\mu0}  &  =\frac{e}{4\pi}\left[  \frac{\partial}{\partial
x_{1}}\left(  \frac{-\gamma\left(  x_{1}-\frac{v}{c}x_{0}\right)  }{W^{3}%
}\right)  +\frac{\partial}{\partial x_{2}}\left(  \frac{-\gamma x_{2}}{W^{3}%
}\right)  +\frac{\partial}{\partial x_{3}}\left(  \frac{-\gamma x_{3}}{W^{3}%
}\right)  \right]  =\nonumber\\
&  =\frac{e\gamma}{4\pi}\left[  \frac{\partial}{\partial x_{i}^{\prime}%
}\left(  \frac{-x_{i}^{\prime}}{r^{\prime3}}\right)  \right]  ,\nonumber\\
\partial^{\mu}F_{\mu1}  &  =\frac{e}{4\pi}\left[  \frac{\partial}{\partial
x_{0}}\left(  \frac{\gamma\left(  x_{1}-\frac{v}{c}x_{0}\right)  }{W^{3}%
}\right)  +\frac{\partial}{\partial x_{2}}\left(  \frac{-\gamma\frac{v}%
{c}x_{2}}{W^{3}}\right)  +\frac{\partial}{\partial x_{3}}\left(  \frac
{-\gamma\frac{v}{c}x_{3}}{W^{3}}\right)  \right]  =\nonumber\\
&  =\frac{v}{c}\frac{e\gamma}{4\pi}\left[  \frac{\partial}{\partial
x_{i}^{\prime}}\left(  \frac{-x_{i}^{\prime}}{r^{\prime3}}\right)  \right]  ,
\label{A2_4}%
\end{align}
The second and third components are%
\[
\partial^{\mu}F_{\mu2,3}=\frac{e\gamma x_{2,3}}{4\pi}\left[  \frac{\partial
}{\partial x_{0}}\left(  \frac{1}{W^{3}}\right)  +\frac{v}{c}\frac{\partial
}{\partial x_{1}}\left(  \frac{1}{W^{3}}\right)  \right]  =\frac{e\gamma
x_{_{2,3}}^{\prime}}{4\pi}\left[  \frac{\partial}{\partial x_{0}}+\frac{v}%
{c}\frac{\partial}{\partial x_{1}}\right]  \left(  \frac{1}{r^{\prime3}%
}\right)  .
\]
This is zero, because\
\begin{equation}
\left[  \frac{\partial}{\partial x_{0}}+\frac{v}{c}\frac{\partial}{\partial
x_{1}}\right]  r^{\prime-3}=\left[  -\gamma\frac{v}{c}\frac{\partial}{\partial
x_{1}^{\prime}}+\gamma\frac{v}{c}\frac{\partial}{\partial x_{1}^{\prime}%
}\right]  r^{\prime-3}\equiv0,\text{ \ }\left[  \frac{\partial r^{\prime-3}%
}{\partial x_{0}^{\prime}}\right]  _{x^{\prime}=const}=0.\nonumber
\end{equation}
Referring to the linear Maxwell equation in the rest frame
\begin{equation}
\frac{\partial}{\partial x_{i}^{\prime}}\left(  \frac{-x_{i}^{\prime}%
}{r^{\prime3}}\right)  =4\pi\delta^{3}(\mathbf{x}^{\prime}), \label{A2_5}%
\end{equation}
we can write%
\begin{equation}
\partial^{\mu}F_{\mu\nu}=4\pi\left(  e\gamma\delta^{3}(\mathbf{x}^{\prime
}),\frac{v}{c}e\gamma\delta^{3}(\mathbf{x}^{\prime}),0,0\right)  .\nonumber
\end{equation}
This coincides with the current (\ref{j''}).

\section*{Appendix 2\label{Ap2}}

In this Appendix we present a detailed derivation of the partial conservation
law Eq. (\ref{partial_conservation}) for a more general Lagrangian than
(\ref{lagrangian})
\begin{equation}
L(x)=-\mathfrak{F}(x)+\mathfrak{L},\text{ \ }\mathfrak{L=L}(\mathfrak{F}(x)),
\label{A1}%
\end{equation}
where $\mathfrak{L}$ is an arbitrary function of $\mathfrak{F}(x)$. The second
pair of nonlinear Maxwell equations in place of (\ref{eq10}) now reads%
\begin{equation}
\frac{\partial}{\partial x^{\nu}}\left[  \left(  1\mathfrak{-}\frac
{\mathfrak{\partial L}}{\partial\mathfrak{F}(x)}\right)  F^{\mu\nu}\right]
=-\frac{1}{c}j^{\mu}, \label{A3}%
\end{equation}
and the stress-energy tensor (\ref{T}) becomes
\begin{equation}
T^{\mu\nu}=-F^{\mu\lambda}F_{\lambda}^{\nu}\left(  1\mathfrak{-}%
\frac{\mathfrak{\partial L}}{\partial\mathfrak{F}(x)}\right)  +g^{\mu\nu
}\mathfrak{F}(x)-g^{\mu\nu}\mathfrak{L.} \label{A4}%
\end{equation}

To obtain the partial continuity equation (\ref{partial_conservation}) one
needs to calculate the derivative $\frac{\partial T^{\mu\nu}}{\partial x^{\mu
}}$ on solutions of the field equations, i.e. by using (\ref{A3}) and the
Bianchi identities (\ref{Bianchi}). We have
\begin{equation}
\frac{\partial}{\partial x^{\mu}}\left[  -\left(  1\mathfrak{-}\frac
{\mathfrak{\partial L}}{\partial\mathfrak{F}(x)}\right)  F^{\mu\lambda}%
F_{\nu\lambda}\right]  =-\frac{1}{c}j^{\lambda}F_{\nu\lambda}-\left(
1\mathfrak{-}\frac{\mathfrak{\partial L}}{\partial\mathfrak{F}(x)}\right)
F^{\mu\lambda}\frac{\partial F_{\nu\lambda}}{\partial x^{\mu}}, \label{A5}%
\end{equation}
where%
\begin{equation}
-\delta_{\nu}^{\mu}\frac{\partial}{\partial x^{\mu}}\left[  -\mathfrak{F(}%
x\mathfrak{)+L}\right]  =\left(  1\mathfrak{-}\frac{\mathfrak{\partial L}%
}{\partial\mathfrak{F}(x)}\right)  \frac{\partial\mathfrak{F(}x\mathfrak{)}%
}{\partial x^{\nu}}. \label{A7}%
\end{equation}
Let us calculate the derivative
\begin{equation}
\frac{\partial\mathfrak{F(}x\mathfrak{)}}{\partial x^{\nu}}=\frac{1}%
{2}F^{\lambda\rho}\frac{\partial F_{\lambda\rho}}{\partial x^{\nu}}.
\label{A8}%
\end{equation}
Next we use (\ref{Bianchi}) to transform the derivative $\frac{\partial
F_{\lambda\rho}}{\partial x^{\nu}}$ as
\[
\frac{\partial F_{\lambda\rho}}{\partial x^{\nu}}=-\frac{\partial F_{\rho\nu}%
}{\partial x^{\lambda}}-\frac{\partial F_{\nu\lambda}}{\partial x^{\rho}}.
\]
Now Eq. (\ref{A7}) can be written as%
\begin{equation}
-\delta_{\nu}^{\mu}\frac{\partial}{\partial x^{\mu}}\left[  -\mathfrak{F(}%
x\mathfrak{)+L}\right]  =\left(  1\mathfrak{-}\frac{\mathfrak{\partial L}%
}{\partial\mathfrak{F}(x)}\right)  \frac{1}{2}F^{\lambda\rho}\left(
-\frac{\partial F_{\rho\nu}}{\partial x^{\lambda}}-\frac{\partial
F_{\nu\lambda}}{\partial x^{\rho}}\right)  . \label{A9}%
\end{equation}
Taking into account that $F_{\mu\nu}$ is antisymmetric, and making the change
of the summation indices $\lambda\rightarrow\mu$ in the first term, and
$\rho\rightarrow\mu$ in the second, we can see that the both terms in the
latter expression coincide and their sum is just the second term in (\ref{A5})
with the opposite sign to cancel its contribution into the full derivative of
$T_{\nu}^{\mu}$. Gathering together (\ref{A5}) and (\ref{A9}) we finally
obtain (\ref{partial_conservation}).

\end{document}